# Navigating the unknown in large-scale operational transformation programs: The “Sirius Days” framework as a ‘pilot-organization’ for characterizing emerging issues

Françoise Zink[1], Chipten Valibhay[1], Jose Bonet Faus[1][2]

(1) Mines Paris, Université PSL, Centre de Gestion Scientifique (CGS), i3 UMR CNRS, 75006 Paris, France
(2) Ubisoft, 2 Avenue Pasteur, 94160, Saint-Mandé, France

**Abstract:** Large-scale digital transformation programs must simultaneously sustain existing operations and navigate deep unknowns emerging from IT-business-operations interactions — a challenge conventional project governance frameworks inadequately address. Based on a longitudinal case study of a transformation program, we investigate the "Sirius Days," a monthly senior management retreat identified as a critical success factor. We show that this framework constitutes a *pilot-organization*: an organizational *‘dispositif’* (or apparatus) that deconstructs established knowledge or assumptions, formulates rigorous conjectures, and tests them in real conditions. It generated five resilience levers —systemic characterization of unknowns, early anomaly discernment, expansion of performance norms, social capital creation through a community of inquiry, and expansion of organizational agility across scales —revealing a model of an organizational ‘dispositif’ that operationalizes navigating unknowns across cognitive, social, and normative dimensions.

## 1. Context: understanding how to navigate unknowns in large-scale digital transformation programs.

Historically, researchers have distinguished project management modalities based on the degree of uncertainties and unknowns – separating exploratory from exploitation project and studying their relations. Large-scale digital transformation programs, related to critical operations such as dealing with legal documents, cash flows, and ensuring on-going operations must simultaneously ensure existing performances of the organization and the exploration and implementation of innovative tools and processes. Quite surprisingly, this state-of-affair echoes many industrial contexts and technical systems, in which major transformations must occur while the system is still operating, calling for specific interventional approach (Maier et al., 2022 ; Jibet, 2025 ; Bonet Faus et al, 2026).

From the academic perspective, Midler (2012) developed an ideal-typical model where uncertainty decreases as the project converges toward production. However, this model may overlook remaining challenges in cases where projects are dealing with radical uncertainties or unknowns. Loch et al. (2006) theorized specific management regimes for such environments. However, the question of adapted governance and methods for large-scale digital transformation programs—and the actual strategic benefits companies derive from their success— remains an open debate (Correani et al., 2020; Wu et al., 2023).

## 2. Literature Review and Research Question: inventing a *'dispositif'* to deal with the transformation of an existing system.

Literature characterizes digital transformation projects not only by their technical nature but also by their extensive interactions with business operations (Barthel & Hess, 2019). This concurrency of design and operations raises several managerial challenges, which are heightened when the system undergoing such a profound transformation is an "engineering system", comprising many constituent systems, products, and services (Maier et al., 2022) (De Weck et al., 2011). This term appropriately describes global infrastructure systems such as energy, transport or water management networks, which are characterized by heavy legacy, dynamic interactions with their environment and long and uncertain life cycles.

In this context, a digitalization project can be conceptually understood as an "intervention" (Maier et al., 2022). Its goal is to improve the performance of a specific process within a larger system of interest by digitalizing the subsystem that most constrains the targeted measure of performance, while minimizing the disruption of contingent processes (Bots, 2022). Given the aforementioned specificities of engineering systems, any such intervention must inevitably deal with unexpected events, manage risks, revise initial assumptions, and extend the scope of relevant stakeholders of the project (Lenfle, 2008).

Sound management of this emergence of unknowns along the intervention process requires two key capabilities:

1. an ambidextrous mastery of "deliberate exploration" and "emerging exploration" (Tillement et al. 2019), meaning a balanced integration of rigorous planning and controlled execution paired with a flexible exploration of alternative paths to acquire new knowledge and adapt to changing conditions (Lenfle et al. 2010).
2. a willingness to acknowledge ignorance, crucial in identifying the nature and degree of unknowns (Hatchuel, 2023).

Despite historical roots, literature points out that predominant approaches in modern engineering and management disciplines do not inherently guarantee the development of these capabilities (Lenfle et al. 2010). Which is supported by empirical evidence highlighting the low success rates of large digital transformation programs (Wu et al., 2023). There thus emerges a need for a "managerial technique" (Weil et al. 1992) capable of ensuring the development and preservation of these capabilities, while preventing the costly pitfalls associated with a failing large scale digital transformation intervention

A recent work (Valibhay & Hatchuel, 2025) concerning contexts where dealing with unknowns is critical suggest that certain organizational conditions are conducive to the development of the aforementioned capabilities. Valibhay and Hatchuel further specify that the genesis of these organizational conditions hinges on the invention of specific organizations dedicated to identifying unknowns and building generative capacities to address them. This type of organization could be analyzed as a "dispositif", roughly translated to apparatus, a concept rooted in French organizational theory (Aggeri & Gand 2025) which describes *the means* by which a collective action is conducted, including techniques, methods, processes, actors.

In short, we would like to show that managing emerging unknowns for large-scale digital transformation programs could require the invention of such "dispositif" dedicated to navigating unknowns. There are relatively few empirical examples describing how such a “dispositif” is designed to be an operational instrument and, to the best of our knowledge, they are marginally addressed. This leads us to the following question: **What are the required features of an organizational “dispositif” dedicated to navigating the emergence of unknowns in the context of large-scale transformation projects?**

## 3. Methodology: an exploratory case study of an organizational invention - “The Sirius Days”.

### 3.1 Description of the case study

This research is based on a longitudinal exploratory case study within a major service company. The large-scale digital transformation project – the "Alpha Program" (€ 80M, 6 years) - aims to transform the Customer Services Domain and is structured into two phases (Front then Back Office). The Backoffice phase, referred to as ‘Program Sirius’, required migrating 4.6 million active customer accounts across 8 legacy regions onto a unified, modernized revenue engine. Managing this transformation required 20 to 30 parallel sub-projects to pay down technical debt (Li et al., 2015) and modularize the legacy ERP (Enterprise Resource Planning).

That phase began under significant historical tension (technical debt, past failures) and was further impacted by major regulatory changes. Although designed in a modular and learning-oriented fashion, in the winter of 2022-2023 the program struggled to address the challenge of scaling up. However, by early 2026, the Alpha program is considered a great success: online implementation was achieved with near-perfect data integrity and billing accuracy, alongside high performance in the sales cycle. According to stakeholders (surveyed in Q4 2025 and Q1 2026), the "Sirius Days" — a monthly retreat format invented by senior management — were a key success factor. This prompted our investigation.

Deep contextual immersion was ensured by a main author serving as the Head of Backoffice from 2018 to March 2025, providing privileged access to strategic insights and continuous observation. Reflective distance was established between April 2025 and April 2026 through firsthand observations of the final two critical cutover phases (October 2025 and April 2026) and ex-post qualitative interviews with core executive participants.

### 3.2 Data collection

The comprehensive corpus of empirical materials collected and analyzed includes:

1. 32 "Sirius Days" Slide Sets: Comprehensive corporate records containing meeting agendas, emerging topics briefs, and formal "so what" logs collected monthly between December 2022 and March 2025.
2. 20 Emerging Topics Investigations: Full documentation from the specialized tracking team containing flash diagnoses, internal cross-functional expert reviews, and comprehensive impact studies.
3. External Shockwave Logs**:** Contextual data regarding exogenous events that disrupted the program's timeline (e.g., the 2023 macroeconomic drought issue, shifting interest rates, and the implementation of new national ‘*Agence de l'Eau’* billing legislation).

These unique data, as well as the immersion from the main author, give us the opportunity to dive into the life of the project and understand how certain unforeseen 'events' were managed and the role played by the Sirius Days.

### 3.3 Data coding

The materials were first reviewed to verify creation dates and the consistency of references between the 20 emerging topic files and the 32 Sirius Days. This analysis was facilitated by the fact that they were created using SaaS tools that maintain a history log and contribution tracking.

After verifying that emerging topics were at the heart of the Sirius Days (systematically on the agenda, except during instruction staff leave periods), we analyzed the entirety of the 'emerging topic' drive (shared drive as of May 25, 2026), excluding 2 topics relating to finance and not to the Sirius program. This drive contains two methodological standards: the flash diagnosis and the impact assessment.

20 topics were listed and encoded using a 7-dimension presence/absence matrix. The first two dimensions verify the methodological self-alignment of the investigations carried out. The second two dimensions aim to assess if these topics forced knowledge expansion (Delta K) and organizational change (Delta R). These two dimensions characterize if these topics are merely managed as marginal adaptation to the initial program's plan or if they triggered a deeper reconfiguration of the program. The following three dimensions focus on the typology of actions implemented after the investigation of the emerging topics and characterize the integration pathways chosen by the Sirius Days collective.

The table below describes, for each dimension, the components of the analysis. The resulting matrix for the 20 emerging topics is provided in Appendix (Table 2).

| **Dimensions** | **Encoding criterion: 1 if yes, 0 if no** |
|---|---|
| Flash Diagnostic | There is a formalized document that complies with the flash diagnostic standard |
| Impact assessment | There is a formalized document that complies with the impact assessment standard |
| Delta K (Knowledge) | There is a formalized document containing a glossary, a map, a framework, a process or part of a process, or a quantification (flows, work units, KPIs, non-compliance) that had not been previously described (acquisition) or that was incomplete or even incorrect (reorganization). |
| Delta R (Organizational Relationships) | There is a formalized document containing a legacy/Sirius comparison or a new Sirius guideline with roles and responsibilities whose boundaries are changing, or which involve new stakeholders. |
| Project included in the Sirius program and delivered | The defined project is integrated into the Sirius program project portfolio and delivered within this framework; it is either individualized in the governance with a recurring project committee, or, if it is smaller in scope, it is embedded in the quarterly releases. |
| Knowledge base update for change, migration, modeling | The knowledge base held by Mission Sirius (change management) has evolved to integrate the glossaries, processes, standards, and frameworks established during the investigation of emerging topics, following validation within the business governance framework. Glossaries and frameworks have been intensively used by the IT/Business teams for contract modeling and technical data migration (cutover). |

| BO (Back Office) domain action plan completed | A business action plan has been defined and completed: for example, regarding the topic of euro-denominated down payments, it was decided to switch to volume-based down payments in legacy systems, rather than developing a specific customization in the Sirius core billing system to handle euro down payments. |
|---|---|

***Table 1 Encoding criterion***

## 4. Empirical Findings

### 4.1 The origin of the Sirius days : inventing a new organization to deal with unknowns in large-scale programs.

The Sirius program spanned from 2021 to 2026, focusing on the transformation of back-office operations within the consumer services domain. Designed from the outset as a multi-year initiative, it progressed through three key stages: a pilot phase, an industrialization phase, and a scaling-up phase. The program was built on principles of modularity—managing a portfolio of 20 to 30 projects—and continuous learning, following a "no-regret" pathway.

Figure 1 (extracted from the Sirius materials) illustrates the governance framework that was implemented.

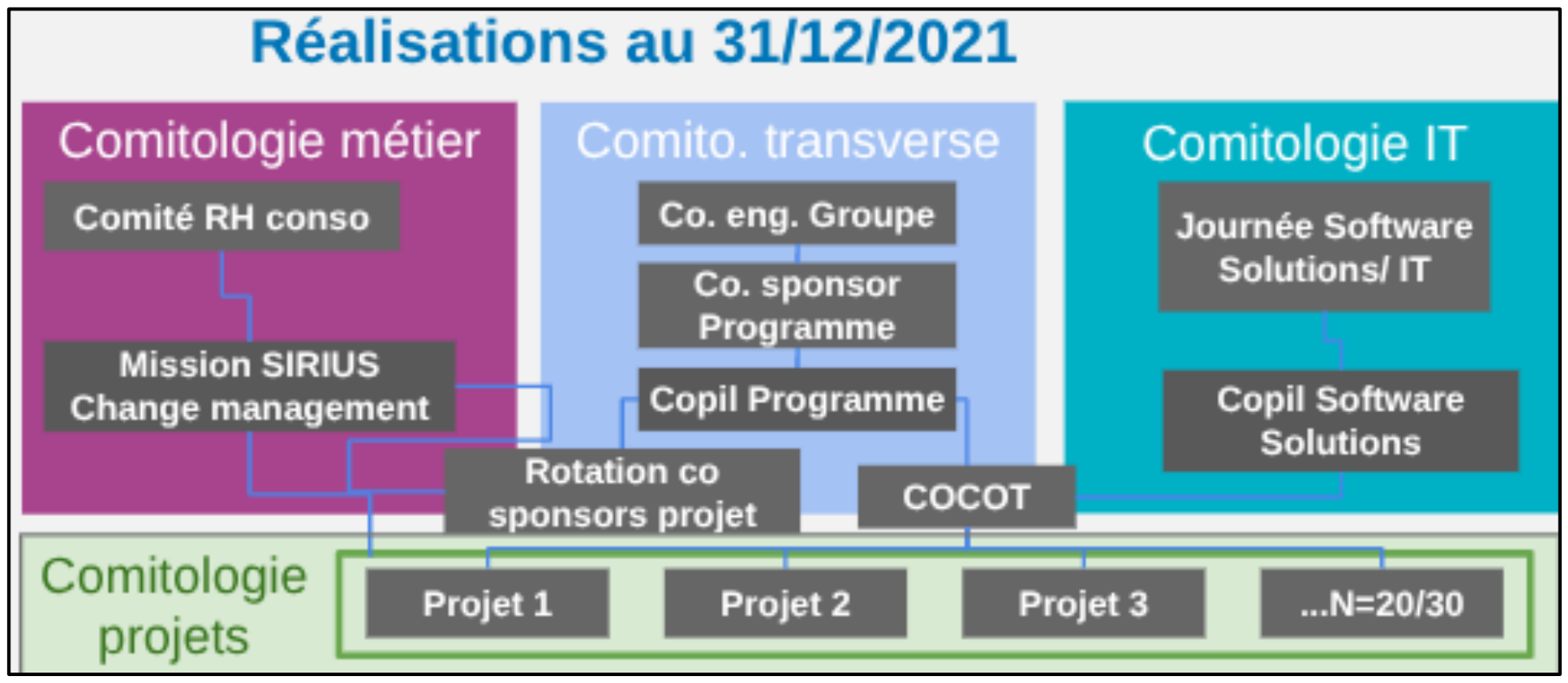

***Figure 1 – Governance framework of the program***

The program was managed using a risk-driven approach based on a standard probability/impact matrix, as illustrated in Figure 2 (extracted from the Sirius Days materials). The risk identification and scoring process followed a two-step approach: first, standard project risk management was conducted at the program control level, where risks were logged and scored by the individual projects and then aggregated by the program. Second, this draft was systematically presented to stakeholders, allowing them to expand the risk register or challenge the scoring based on their own values.

In addition, the following elements can be noted:

- A three-pillar change management approach: driving learning through the Knowledge Base, providing hands-on coaching and support, and delivering clear communications.
- Agile HR feedback loops: implementing regular employee pulse surveys to monitor organizational health.
- Unwavering sponsorship at both the country level (France CEO) and the corporate level (Group CFO).
- Stakeholder onboarding processes.

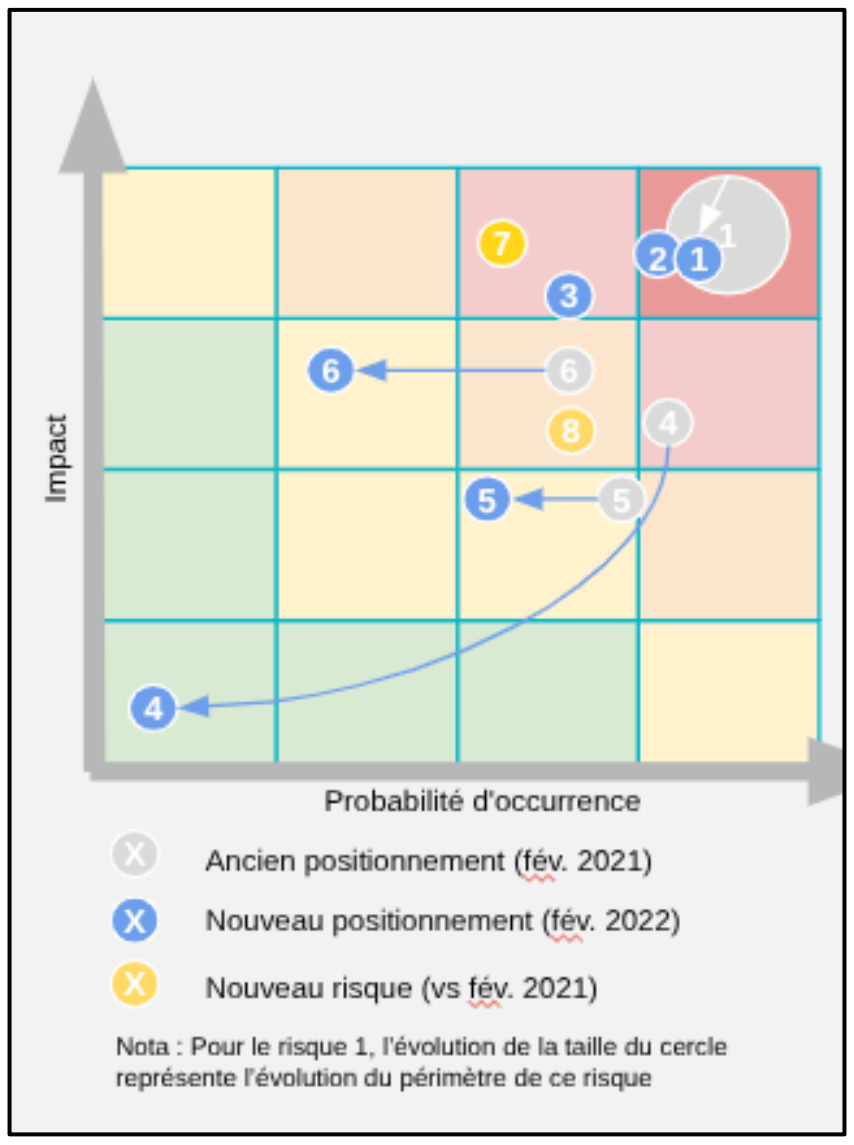

***Figure 2 – Probability/impact matrix for risk assessment and follow-up***

The program was therefore consistent with the state-of-the-art principles described in the academic literature on the governance and control of large scale transformations (see : Sanchez, R. and Mahoney, J.T. (1996) on modularity ; Midler, Lenfle on project and program management ; Oehmen, J., & Vermaas, P. E. 2022 on explicit management practices of uncertainties ; Kotter, J. P. (1995) on change management).

The Sirius Days initially emerged as an empirical response to the unknowns associated with scaling-up and the mechanism remained in place as long as unresolved questions continued to exist regarding large-scale deployment, as illustrated in Figure 3 below and in the log of the 32 Sirius Days held between Q3 2022 and Q1 2025.

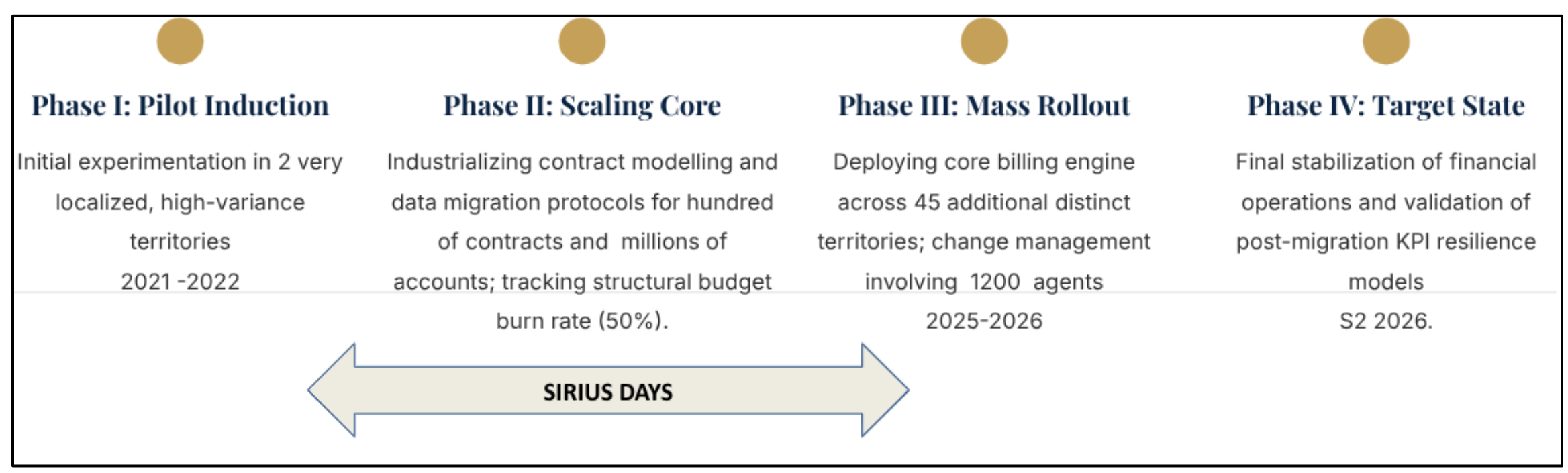

***Figure 3 - Program timeline with phases.***

The following testimony captures well how the Sirius Days were perceived by its participants:

> *'This morning, I was asked for my feedback on SIRIUS, and I found it difficult to put a label on these Sirius Days. I described them as days for alignment, for clearing the air, and for allowing disagreement while always seeking solutions—a sort of "cementing room" for the core team. (...) Ultimately, it was about regularly ensuring that we were all in the same boat.'* (GDC email, October 8, 2025).

Empirical data reveals that the Sirius Days were structured around four key pillars: an executive hybrid core (4–5 leaders across Business, IT, and Program management with full EXCOM budget and delegated authority); a dedicated agile staff (1–3 business-facing analysts); recurring periods of reflective isolation and rigorous exploration methods.

### 4.2 Navigating the unknowns: impacts of the Sirius Days

The 20 emerging topics investigated during the longitudinal study were tracked across their operational outcomes (see Table 3).

Knowledge is demonstrably central to the lifecycle of these emerging topics: their exploration generated new knowledge in 85% of cases, while 75% of the resulting operational outcomes focused on knowledge capitalization and sharing. Organizational relationships represent another critical dimension: 60% of emerging topics identified a need for change, characterized by boundary shifts driven by new processes, stakeholders, or dependencies. Furthermore, 50% of the selected operational outcomes aimed at the adoption of new domain standards, interactions, and rituals.

| **Dimension evaluated** | **Implication Rate (%)** |
|---|---|
| Delta K (Knowledge / Competences) | **85%** |
| Delta R (Organizational Relationships) | **60%** |
| Integrated and Delivered Project within Sirius | **30%** |
| Knowledge base update; Change management (Mission Sirius); Contract modeling | **75%** |
| Executed Back Office (BO) Action Plan | **50%** |

***Table 3 - Assessment of the coded dimensions across the 20 topics.***

Only 30% of the topics culminate in an integrated project delivered within the Sirius program framework. Direct integration is selected for critical workstreams (e.g., the accounting core, document management, and technical debt resolution).

The distribution of chosen operational outcomes provides ex-post evidence of the framework's generative capacity, demonstrating organizational impacts of this ‘dispositif’.

### 4.3 The structure and methods of the Sirius Days

#### 4.3.1 Activities

The main recurring activities of the sessions were as follows.

- Emerging topics involved the detailed examination of one or several issues depending on the progress achieved.
- Addressing these topics that generated significant friction (or for which friction was anticipated) and whose resolution did not fall solely within the responsibility of either the business units or the program. These moments also provided opportunities for participants beyond the core team to temporarily join the group during the workshop.
- In both activities discussions focused on the “So what?” question, insisting on assessing impacts and potential solution paths *without taking decisions* (suspended decision-making).

Few features must be highlighted: first, these “emerging topics” are not self-evident and require in-depth analysis to name and understand what type of unknowns they covered. Therefore, these activities aim to ‘reveal’ unknowns and structure them so that exploration could occur. Quite surprisingly, it was not a “task force” dedicated to problem-solving, nor an executive committee dedicated to allocating resources on specific challenges. Second, these activities aim primarily to ‘socialize’ the unknowns and study them, and ensure that they are recognized as such, without necessary taking decisions – except decisions of further inquiry.

#### 4.3.2 Methods

The objective was to assess the nature of the emerging issues: “What was actually going on here?”. In many cases, the investigation team had to challenge the assumptions underlying long-established practices, including those held by “domain experts”, and frequently concluded that “this did not align with the facts”.

The methods used to explore emerging topics were codified during the winter of 2022–2023 in the form of a capture mechanism and two standards.

**The capture mechanism** cross-referenced qualitative inputs (Gemba walks, barometers, coffee chats, workshops, insights from Steerco or previous Sirius Days), anomaly reporting (customer complaints, non-conformities, SLAs, IT/IST Run ticket logs, and Business/IT Run post-mortem reviews), KPI deviations, as well as continuous competitive and regulatory monitoring.

**The flash diagnosis** included an impact matrix covering processes, tools, and organization, and provided an action template aligned with the company’s strategic priorities (e.g., offer competitiveness, cash, and customer satisfaction). Figure 4 provides an illustration of a flash diagnosis extracted from the Sirius materials, the advanced payments in Euros.

If the decision was made to pursue further investigation, **an impact study** was conducted using a risk and affected-domain matrix. The study produced several key outputs, including the establishment of a shared glossary, as well as a detailed process flow analysis comparing pre- and post-migration scenarios. It also included a risk assessment of maintaining the status quo by evaluating the potential consequences of taking no action.

The “Flash Diagnosis” was initiated in 65% of cases and served as the primary entry point for rapidly qualifying anomalies or strategic opportunities.

The “Impact Study” followed in 55% of instances, indicating that more than half of the identified topics presented structural complexities requiring in-depth analysis.

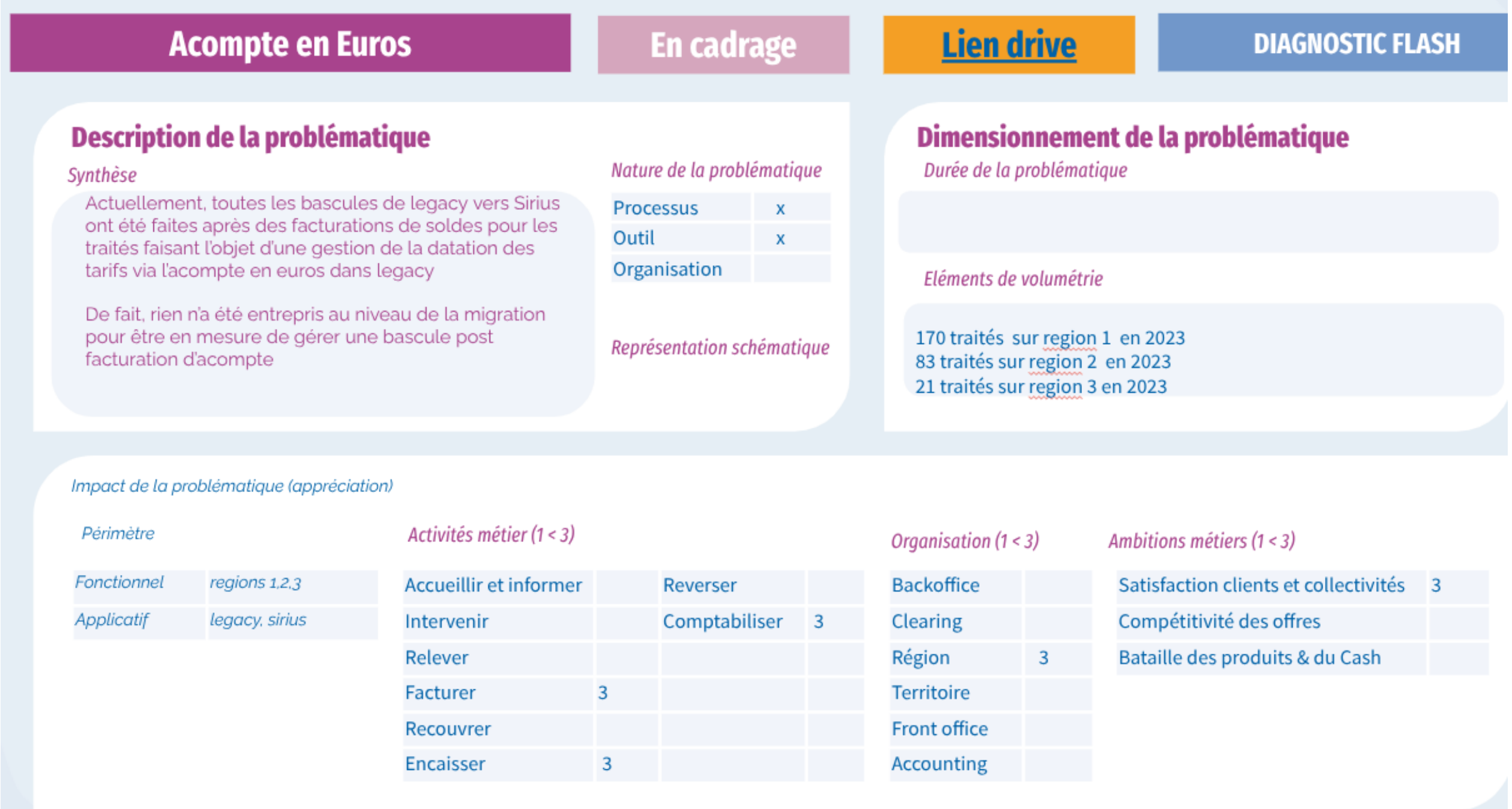

*Figure 4. Illustration of ‘flash diagnosis’ template.*

### 4.3.3 Key features of the ‘Sirius Days’

The Sirius Days used five levers or mechanisms enhancing socio-organisational resilience (Wied et all 2020, Oehmen et Kwakkel, 2022):

- **Early Identification of Emerging Topics:** Identifying critical issues before they arose directly populated a dedicated prioritization backlog. This process relied on gathering open questions or anomalies from both the program and business units through multiple channels and data streams.
- **Systemic Characterization of Unknowns:** Weighing issues precisely allowed for informed decisions on whether to explore them, integrate them into the program or business, or close them entirely.
- **Expansion of Performance Norms:** Redefining success strategically shifted the focus from a complex "switchover" to the capacity to "operate the full annual sales cycle to the day," thereby expanding the program's strategic boundaries.
- **Building Social Capital and Trust through a Community of Inquiry:** Establishing a safe space, through a “retreat”, enabled collective bets despite strong headwinds. At the executive level, managing ignorance fostered cohesion by nurturing this community of inquiry and streamlining 'So What?' debates to build consensus on shared action models.
- **Expanding Organizational Agility across Scales:** Developing advanced expertise in navigating varying operational scales, alongside the institutionalization of the exploration ritual, enhanced the capacity to orchestrate strategic actions within an environment subject to sudden, severe, and unpredictable scale distortions.

These levers reveal a need to manage unknowns mixing cognitive, social and normative dimensions. While this type of organization has been studied for exploratory or innovative projects, we show that it was a key mechanism to anticipate, absorb and deal with unknown events (and not only uncertainties) during the program.

## 5 Results and Contributions

We frame three main contributions regarding the digital transformation of ongoing operations.

First, we identify the need for ignorance management by senior domain leaders. According to Hatchuel and Weil, 1992, we understand The Sirius Days as a managerial technique to rationalize the navigation in the unknowns of the digital transformation of ongoing operations.

| | |
|---|---|
| **Technical substrate** | Navigation in unknowns spaces using diagnosis, impact assessment, and collective discussion in a safe space. |
| **Managerial philosophy** | Rationalisation of Ignorance<br>Open questions exploration and assessment to discern which question is important enough to trigger action |
| **Simplified vision of organizational relationships** | Community of inquiry, “emerging topics” staff |

***Table 4. Description of the managerial technique used to rationalize ‘ignorance’ and unknowns.***

Second, we highlight the generativity of ignorance management. By mapping and instrumenting spaces to observe weak signals and anomalies, and by tracking and weighting unknowns, the "Sirius Days" generated new or restructured knowledge alongside a continuous stream of open questions. Through this pilot structure, operational executives developed socio-organisational resilience, and a collective capability to reframe the program, business processes, organizational design, and performance standards.

Third, we show that the empirical "Sirius Days" *dispositif* follows a model that we define as ‘pilot-organization’. This model mixes unexpected organizational conditions required to deal with unknowns while having a holistic view on the project and getting out of the project’s momentum. The Sirius Days functioned as a space for deconstructing established knowledge and for the rigorous formulation of logical and quantified conjectures that were then tested and experimented ‘live’. This activity is therefore ‘piloting’ the project’s emerging issues, which is why we coined this organizational invention as ‘pilot-organization’.

## Implications

Practitioners of large-scale digital transformation programs will find in this research a suggested organizational structure between business domains and program that can help—or so we hope—to succeed in both the "build" phase and the "run" of operations during and after the program. This structure could also offer an extension to traditional project governance by adding a 'pilot-organization' as a critical space for exploring unsolved questions and build both cognitive and relational capabilities to address them.

From an academic perspective, the mechanisms emphasized in this study echo mechanisms and institutional structure studied to deal with unknowns at higher scale (ecosystems level, industrial/multi-organizational level or even academic disciplines) (Agogué and al., 2013; Cogez and al., 2013; Valibhay & Hatchuel, 2025). This could suggest further investigation on how this 'management of unknown' could be implemented in the transformation of systems – knowing how critical it is to deal with contemporary grand challenges (Fritzsche and al., 2026).

# Appendix

| no | **Sujet émergent** | Diagnostic flash | Etude d'impact | Delta K (savoirs) | Delta R (relations organisationnelles) | Projet inclus dans programme Siirus et délivré | Màj de la KB pour change, migration, modelisation | Plan d'action domaine BO réalisé |
|---|---|---|---|---|---|---|---|---|
| 1 | Moyens de paiement | 1 | 1 | 1 | 1 | 0 | 1 | 1 |
| 2 | Coefficients d'actualisation | 1 | 1 | 1 | 1 | 1 | 1 | 0 |
| 3 | Gestion des primes fixes | 1 | 1 | 1 | 0 | 0 | 1 | 1 |
| 4 | Vente et cession interne | 1 | 1 | 1 | 1 | 0 | 1 | 0 |
| 5 | Comptabilisation des tripartites | 1 | 1 | 1 | 1 | 0 | 1 | 0 |
| 6 | Structure et gestion des comptes d'attentes | 1 | 1 | 1 | 0 | 0 | 1 | 0 |
| 7 | Organisation nationale de la gestion des redevances | 0 | 1 | 1 | 1 | 1 | 1 | 0 |
| 8 | Pilotage des volumes de facturation en campagne d'hiver 2023 | 0 | 0 | 1 | 1 | 0 | 0 | 1 |
| 9 | Petites évolutions du coeur de facturation, transfert de gouvernance | 0 | 0 | 0 | 1 | 0 | 0 | 1 |
| 10 | Editique facture, paiement de la dette technique | 1 | 0 | 1 | 1 | 1 | 1 | 0 |
| 11 | Cartographie du cycle des ventes | 0 | 0 | 1 | 1 | 1 | 1 | 1 |
| 12 | Recouvrement en territoire en mode Sirius | 0 | 0 | 0 | 0 | 0 | 0 | 0 |
| 13 | Sirius coeur de facturation les grands concepts | 0 | 0 | 0 | 0 | 0 | 0 | 0 |
| 14 | Gestion des Fonds de Solidarité Logement | 0 | 1 | 1 | 0 | 1 | 1 | 0 |
| 15 | Maîtrise de business object | 1 | 0 | 1 | 1 | 0 | 1 | 1 |
| 16 | Editique : sécurisation du référentiel de marques | *1* | *1* | 1 | 1 | 0 | 1 | 1 |
| 17 | Facturation des inter sociétés du groupe | 1 | 1 | 1 | 1 | 1 | 1 | 1 |
| 18 | Comptes de gestion | 1 | 0 | 1 | 0 | 0 | 1 | 1 |
| 19 | Gestion de la mensualisation | 1 | 1 | 1 | 0 | 0 | 1 | 0 |
| 20 | Acomptes en euros | 1 | 0 | 1 | 0 | 0 | 0 | 1 |
| | Total | 13 | 11 | 17 | 12 | 6 | 15 | 10 |
| | % | 65% | 55% | 85% | 60% | 30% | 75% | 50% |

**Table 2 – Encoded dimension for each emerging topic.**